\documentclass[twoside]{dis09}
\usepackage[latin1]{inputenc}
\usepackage[dvips]{graphicx,epsfig,color}
\usepackage{wrapfig,rotating}
\usepackage{amssymb,amsmath,array}

\begin{document}
\title{Path Integral Methods for Soft Gluon Resummation}

%***********************************************************************
% AUTHORS INFORMATION AREA
%***********************************************************************
\author{Chris D. White
%
% Optional short acknowledgment: remove next line if non-needed
\thanks{Talk given at ``DIS 2009", Madrid, Spain, 26-30 Apr 2009 }
%
% DO NOT MODIFY THE FOLLOWING '\vspace' ARGUMENT
\vspace{.3cm}\\
%
% Addresses and institutions (remove "1- " in case of a single institution)
Nikhef, Kruislaan 409, 1098SJ Amsterdam - The Netherlands\\
}
%***********************************************************************
% END OF AUTHORS INFORMATION AREA
%***********************************************************************

\maketitle

\begin{abstract}
We describe a new framework for soft gluon resummation, based on path integral methods. 
The approach recovers previous results obtained in the eikonal approximation, in which all
emitted gluons become soft. Furthermore, a clear physical interpretation allows straightforward
extension of the framework to beyond the eikonal approximation, and we use this to classify the 
structure of next-to-eikonal corrections to matrix elements.
\end{abstract}

\section{Introduction}
The multiple emission of soft gauge bosons leads to large corrections to scattering cross-sections. Typically, if $\xi$ if the energy carried by the gauge bosons, a given cross-section has contributions of schematic form
    \begin{equation}
      \frac{d\sigma}{d\xi}=\sum_{n,m}\alpha^n\left[c^0_{nm}\frac{\log^m(\xi)}{\xi}+c^{1}_{nm}\log^m(\xi)+\ldots\right],
      \label{softstruc}
    \end{equation}
where $\alpha$ is the coupling constant of the gauge theory, and $c_{nm}^i$ are constants. The first set of terms in eq.~(\ref{softstruc}) arises from the {\em eikonal} approximation, in which the 4-momenta of all emissions $k_i^\mu\rightarrow0$. The second set of terms, suppressed by the energy variable $\xi$, corresponds to the next-to-eikonal approximation in which each boson momentum appears linearly in the total expression for the scattering amplitude. It is well known that soft gauge boson corrections up to NE order can have a sizable effect on predicted cross-section calculations, and that in the region of phase space in which $\xi\rightarrow0$, it becomes necessary to resum the problem logarithms. How to do this in the eikonal approximation is known in both abelian and non-abelian gauge theories. In the abelian case, the amplitude ${\cal A}$ for a given hard interaction ${\cal A}_0$ dressed by any number of soft photons is given by~\cite{Yennie:1961ad}
\begin{wrapfigure}{r}{0.5\columnwidth}
\centerline{\includegraphics[width=0.45\columnwidth]{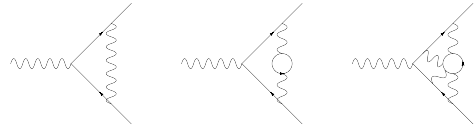}}
\caption{Example diagrams which exponentiate in abelian field theory.}\label{abdiags}
\end{wrapfigure}
\begin{equation}
{\cal A}={\cal A}_0\exp\left[\sum G_c\right],
\label{abexp}
\end{equation}
where the exponent contains connected subdiagrams which span the external particle lines (see figure~\ref{abdiags}).
\begin{wrapfigure}{r}{0.5\columnwidth}
\centerline{\includegraphics[width=0.45\columnwidth]{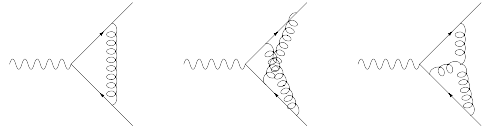}}
\caption{Example webs i.e. diagrams which exponentiate in nonabelian field theory.}\label{nonabdiags}
\end{wrapfigure}
Exponentiation also holds in non-abelian theories, but with a more complicated structure. The equivalent of eq.~(\ref{softstruc}) is~\cite{Gatheral:1983cz,Frenkel:1984pz}
\begin{equation}
{\cal A}={\cal A}_0\exp\left[\sum \bar{C}_W W\right],
\label{nonabexp}
\end{equation}
where the set of subdiagrams $\{W\}$ are two-eikonal line irreducible. That is, they can only be disconnected by cutting the hard external lines more than twice. These diagrams are known as {\em webs}, and examples are shown in figure~\ref{nonabdiags}. The factors $\bar{C}_W$ in eq.~(\ref{nonabexp}) can be interpreted as modified colour factors for the webs, which are indeed different to the normal colour factors one obtains from perturbation theory. 

Having seen the structure of soft gluon corrections at eikonal level, it is natural to investigate the extension of the above ideas to next-to-eikonal order.
\section{Path integral formulation}
\begin{wrapfigure}{l}{0.5\columnwidth}
\centerline{\includegraphics[width=0.3\columnwidth]{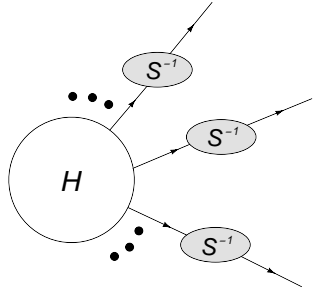}}
\caption{Factorised structure of the Green's function for production of $n$ hard particles dressed by soft gauge bosons (c.f. eq.~(\ref{Green})).}\label{Greenfig}
\end{wrapfigure}
The structure of abelian exponentiation (in terms of connected subdiagrams) is reminiscent of a textbook property of quantum field theory - that disconnected diagrams exponentiate. One may contemplate whether or not these concepts are related, and indeed this turns out to be the case.

We begin with the simpler case of abelian gauge boson emission from scalar particles, and our starting point is the Green's function for production of $n$ hard emitting particles, which may in turn emit soft photons. After separating the gauge field into hard and soft modes, the Green's function has the factorised form~\cite{Laenen:2008gt}
\begin{align}
G(p_1,\ldots p_n)&=\int{\cal D}A_s^\mu H(x_1,\ldots x_n)\notag\\
&\times S^{-1}(p_1,x_1)\ldots S^{-1}(p_n,x_n),
\label{Green}
\end{align}
where the hard interaction $H(x_1,\ldots x_n)$ creates outgoing hard particles at positions $\{x_i\}$, and $S^{-1}(x_i,p_i)$ are propagators for the emitting particles in the background of a soft gauge field sandwiched between states of given initial position and final momentum (i.e. $S$ itself is the quadratic operator in the Lagrangian for the soft gauge field). The path integral in eq~(\ref{Green}) is over the soft gauge field, and the structure is shown schematically in figure~\ref{Greenfig}. 

The propagator for a hard emitting particle in the presence of a soft gauge field can be expressed as a first-quantised path integral (i.e. a quantum mechanics path integral, rather than a quantum field theory one)~\cite{Strassler:1992zr,vanHolten:1995ds}. To do this, one introduces the evolution operator (in time $t$) $U(t)=\exp[-i\frac{1}{2} St]$. Sandwiched between states of given initial position $x_i$ and final momentum $p_f$, this can be written 
\begin{align}
<p_f|U(T)|x_i>&=\int_{x(0)=x_i}^{p(T)=p_f}{\cal D}p\,{\cal D}x\\
&\quad\exp\left[-ip(T)x(T)+i\int_0^T dt (p\dot{x}-{\cal H}(p,x))\right],
\label{pathint}
\end{align}
where the path integral is over all possible phase space trajectories consistent with the boundary conditions, and ${\cal H}$ the Hamiltonian coupling the scalar particle to the soft gauge field. Inserting this into eq.~(\ref{Green}) for each external line, one may perform the integrals over momenta with the result
\begin{align}
G(p_1,\ldots, p_n)&=\int {\cal D}A_s^\mu H(x_1,\ldots x_n)\prod_x e^{-ip_f\cdot x_i}\int {\cal D}x\notag\\
&\quad\times\exp\left[\int_0^\infty dt \left(\frac12 \dot{x}^2+(p_f+\dot{x})\cdot A(x_i+p_f t+x(t))+\frac i2\partial\cdot A(x_i+p_f t+x)\right)\right],
\label{Gres}
\end{align}
where $x_i$ and $p_f$ are the initial position and final momentum of line $x$. The physical interpretation of this result is as follows. Firstly, the path integral is over all possible worldlines of the external particles (i.e. denoted by $x$). Secondly, upon performing the path integrals over these worldlines, eq.~(\ref{Gres}) has the form of a generating functional of a quantum field theory for the soft gauge field. The terms in the exponents generate sources for the gauge field, which are localised along the hard external lines. The path integral over the soft gauge field will generate all possible soft photon subdiagrams spanning the external lines. The exponentiation of soft photon corrections now follows directly from the exponentiation of disconnected diagrams in the quantum field theory whose generating functional is given by eq.~(\ref{Gres}). 

Here I have described the simplest case of abelian exponentiation with scalar emitting particles. The above discussion is easily generalised to the case of fermionic emittors, which merely give an extra term in each exponent of eq.~(\ref{Gres}), corresponding to the fact that beyond the eikonal approximation, emitted particles are sensitive to the spin of the emittor. 

The case of non-abelian gauge theories is more difficult. The above argument no longer applies due to the fact that the source vertices coupling the emitted gauge field to the external lines are matrix-valued in colour space. Thus, the sources do not commute and the usual combinatorics of the path integral that guarantee exponentiation of disconnected diagrams no longer apply. One may make progress, however, using the {\em replica trick} of statistical physics (see e.g. \cite{Replica}). To apply this in the present context, one starts by considering a theory with $N$ identical copies of the soft gauge field $A^\mu_s$, each of which does not interact with the others. The Green's function raised to the power $N$ is given by a simple mathematical identity as
\begin{equation}
G^N=1+N\log{G}+{\cal O}(N^2).
\label{expform}
\end{equation}
One may then examine the various diagrams in the replicated gauge theory (each of which spans the hard external lines as before). Each diagram is proportional to some power of the replica number $N$. Crucially, only a subset of diagrams is linear in $N$, such that from eq.~(\ref{expform}) one has
\begin{equation}
G=G_0\exp\left[\sum\bar{C}_W W\right],
\label{Gsol}
\end{equation}
where $\{W\}$ are those diagrams which are linear in the replica number, and $\{\bar{C}_W\}$ accompanying constants. In~\cite{Laenen:2008gt} the simple setup of two eikonal lines coupled by a colour singlet hard interaction is considered, and the resulting subdiagrams $W$ are found to be the two-eikonal line irreducible webs of~\cite{Gatheral:1983cz,Frenkel:1984pz} (see eq.~(\ref{nonabexp})). Furthermore, the constants $\bar{C}_W$ are precisely the modified colour factors found previously.

\begin{wrapfigure}{r}{0.5\columnwidth}
\centerline{\includegraphics[width=0.3\columnwidth]{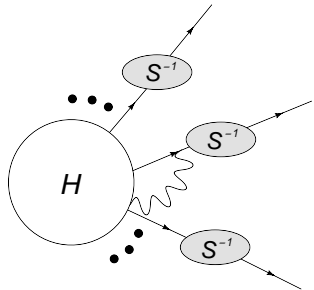}}
\caption{Example of an internal emission graph, contributing to the remainder term in eq.~(\ref{struc}).}\label{internal}
\end{wrapfigure}

The simple physical picture offered by the path integral approach allows straightforward extension to beyond the eikonal approximation, and the basic idea is as follows. The eikonal approximation corresponds to neglecting the recoil of the emitting particles. The path integrals in $x$ of eq.~(\ref{Gres}) can thus be carried out by systematically expanding about the classical straightline trajectories of the emitting scalars. The first set of subleading corrections corresponds to the next-to-eikonal limit~\cite{Laenen:2008gt}, and generates effective Feynman rules for subeikonal emissions which generalise the known Feynman rules of eikonal perturbation theory. Importantly, the arguments for exponentiation given above are unaffected by whether or not one is at eikonal order. Thus, the soft photon (and indeed gluon) corrections from external emission diagrams (i.e. where the soft particles are completely external to the hard interaction) immediately exponentiate, where the next-to-eikonal terms in the exponent consist of a new set of webs involving the NE Feynman rules. This is not the whole story at NE order - there are also subleading contributions from internal emission graphs, in which a genuinely soft ($k^\mu\rightarrow0$) boson is emitted from an external line, but lands inside the hard interaction as shown in figure~\ref{internal}.

\section{Structure of matrix elements up to NE order}
The contribution from internal emissions are fixed by gauge invariance, as described in~\cite{Laenen:2008gt}. This is essentially a rederivation of the well-known Low-Burnett-Kroll theorem~\cite{Low:1958sn,Burnett:1967km}, which relates amplitudes with an extra soft emission to the original amplitude with no such emission. In fact, one has in momentum space (in the abelian case, with a simple modification for the non-abelian case described above)
\begin{equation}
H^\mu(p_1,\ldots p_n)=-\sum_i q_i\frac{\partial}{\partial p_i^\mu} H(p_1,\ldots p_n),
\label{low}
\end{equation}
where $H^\mu(p_1,\ldots p_n)$ is the hard interaction with an extra soft emission, and $q_i$ the charge of the external line $i$. This implies that contributions from internal emission graphs do not exponentiate, but rather have an iterative structure to all orders of perturbation theory. The complete structure of NE corrections to matrix elements is then
\begin{equation}
{\cal M}={\cal M}_0\exp\left[{\cal M}^E+{\cal M}^{NE}\right]\times\left[1+{\cal M}_{rem.}\right]+{\cal O}(NNE),
\label{struc}
\end{equation}
where ${\cal M}_0$ is the Born amplitude, and the remainder term ${\cal M}_{rem.}$ comes from internal emission graphs. The exponent collects the contributions from external emission graphs i.e. webs up to NE order.
\section{Conclusion}
We have introduced a new framework for examining soft gauge boson emission. It uses path integral methods to relate the exponentiation of soft photon and gluon corrections to the known exponentiation of disconnected diagrams in quantum field theory. Old results in the eikonal approximation are recovered, and a classification of next-to-eikonal contributions to matrix elements achieved.
\section*{Acknowledgments}
This work was done in collaboration with Eric Laenen and Gerben Stavenga, and was supported by the Dutch Foundation for Fundamental Matter Research (FOM) and the Netherlands Organisation for Scientific Research (NWO). 

\begin{footnotesize}

%\bibliographystyle{unsrt}
%\bibliography{white_chris.bib}

\begin{thebibliography}{1}
\bibitem{url} Slides: \\ 
\verb$http://indico.cern.ch/contributionDisplay.py?contribId=268&sessionId=3&confId=53294$

\bibitem{Yennie:1961ad}
D.~R. Yennie, Steven~C. Frautschi, and H.~Suura.
\newblock {The infrared divergence phenomena and high-energy processes}.
\newblock {\em Ann. Phys.}, 13:379--452, 1961.

\bibitem{Gatheral:1983cz}
J.~G.~M. Gatheral.
\newblock {Exponentiation of eikonal cross-sections in nonabelian gauge
  theories}.
\newblock {\em Phys. Lett.}, B133:90, 1983.

\bibitem{Frenkel:1984pz}
J.~Frenkel and J.~C. Taylor.
\newblock {Nonabelian eikonal exponentiation}.
\newblock {\em Nucl. Phys.}, B246:231, 1984.

\bibitem{Laenen:2008gt}
Eric Laenen, Gerben Stavenga, and Chris~D. White.
\newblock {Path integral approach to eikonal and next-to-eikonal
  exponentiation}.
\newblock {\em JHEP}, 03:054, 2009.

\bibitem{Strassler:1992zr}
Matthew~J. Strassler.
\newblock {Field theory without Feynman diagrams: One loop effective actions}.
\newblock {\em Nucl. Phys.}, B385:145--184, 1992.

\bibitem{vanHolten:1995ds}
J.~W. van Holten.
\newblock {Propagators and path integrals}.
\newblock {\em Nucl. Phys.}, B457:375--407, 1995.

\bibitem{Replica}
M~Mezard, G~Parisi, and M~Virasoro.
\newblock {Spin Glass Theory and Beyond}.
\newblock World Scientific (1987) 476pp.

\bibitem{Low:1958sn}
F.~E. Low.
\newblock {Bremsstrahlung of very low-energy quanta in elementary particle
  collisions}.
\newblock {\em Phys. Rev.}, 110:974--977, 1958.

\bibitem{Burnett:1967km}
T.~H. Burnett and Norman~M. Kroll.
\newblock {Extension of the low soft photon theorem}.
\newblock {\em Phys. Rev. Lett.}, 20:86, 1968.

\end{thebibliography}
%
\end{footnotesize}

% ****************************************************************************
% END OF BIBLIOGRAPHY AREA
% ****************************************************************************

\end{document}